\documentstyle[12pt,aaspp4]{article}

\def\IRAS{{\it IRAS\/}}
\def\micron{$\mu$m}

\def\deg{$^\circ$}

\def\sevbysev{$7^{\circ}\times 7^{\circ}$}
\def\roph{$\rho$ Ophiuchus}

\begin{document}

\title{The High Resolution {\IRAS} Galaxy Atlas}
\author{Yu Cao}
\affil {Division of Physics, Mathematics, and Astronomy, 
California~Institute~of~Technology, Pasadena, California~91125}
\authoremail{yucao@srl.caltech.edu}
\author{Susan Terebey}
\affil
{Extrasolar Research Corporation, 
Pasadena,~California~91106}
\authoremail{st@extrasolar.com}
\author{Thomas A. Prince}
\affil {Division of Physics, Mathematics, and Astronomy, 
California~Institute~of~Technology, Pasadena, California~91125}
\authoremail{prince@srl.caltech.edu}
\author{Charles A. Beichman}
\affil
{Infrared Processing and Analysis Center,~California~Institute~of~Technology,
Pasadena,~California~91125}
\authoremail{chas@ipac.caltech.edu}

\received{16 October 1996}
\accepted{25 March 1997}

\begin{abstract}
An atlas of the Galactic plane
($-4.7${\deg} $< b <$ 4.7{\deg}) plus the molecular clouds
in Orion, {\roph}, and Taurus-Auriga has been produced at 60 and 100 {\micron}
from {\IRAS} data. The atlas consists of resolution-enhanced coadded images
having $1'$ -- $2'$ resolution as well as coadded images at the native {\IRAS}
resolution. The {\IRAS} Galaxy Atlas, together
with the DRAO HI line / 21 cm continuum and FCRAO CO (1-0) line Galactic
plane surveys, both with similar ($\sim$ 1{\arcmin}) resolution,
provide a powerful venue for studying the interstellar medium, star
formation and
large scale structure in our Galaxy. This paper documents the production and
characteristics of the Atlas.

\end{abstract}


\section{Introduction}
In 1983 the {\em Infrared Astronomical Satellite} fundamentally
changed our view of the infrared sky when it conducted the first infrared
all sky survey. The {\IRAS} data have proven important to the study
of many astrophysical phenomena including star formation, the
interstellar medium, Galactic structure, late-type stars, supernova remnants,
external galaxies, infrared cirrus, and debris disks around nearby stars
(\cite{Beichman87}).
Newer infrared spacecraft missions such as ISO, MSX, and IRTS now provide 
higher sensitivity and spatial resolution (\cite{ISO95,IRTS95,MSX95,MSX96}). 
However, by design they cover only a small fraction of
the sky, thus ensuring the {\IRAS} data will provide a fundamental archive for many years to come.

The native spatial resolution of the {\IRAS} coadded data
is a few by five arcminutes. Various image reconstruction
techniques have been applied to the {\IRAS} data in the quest
to extract higher spatial resolution (\cite {Terebey94}).
These include maximum entropy techniques, among them 
the HIRAS package developed at Groningen
(\cite{Assen95,Bont94}). Making use of an alternate approach, 
the production of the {\IRAS} Galaxy Atlas 
is based on the well-known HIRES processor, 
first developed in 1991 and made available
to the scientific community by
the Infrared Processing and Analysis Center (IPAC). 
HIRES implements the iterative
Maximum Correlation Method (MCM; \cite {MCM90}), a variant
of the Richardson-Lucy algorithm which has been optimized for
{\IRAS} data. The advantages of HIRES include
flux conservation, speed of processing, and the ability to work reliably on 
faint sources.  HIRES images have been
successfully used for a variety of Galactic and extragalactic studies
(\cite{Rice93,Surace93,Terebey94}).

The parallel supercomputing facilities available at Caltech and
the development of new artifact reduction algorithms made possible
a large-scale high-resolution {\IRAS} mapping of the Galactic plane 
(\cite{Cao96}, hereafter referred to as Paper~I, \cite{Cao97}).
The new {\IRAS} Galaxy Atlas (IGA) maps have $1'$ -- 2 $'$ resolution. This 
represents a three-fold improvement in linear resolution for a total factor of ten 
improvement in areal resolution over 
the {\IRAS} Sky Survey Atlas (ISSA; \cite {ISSA94}).
The IGA incorporates several important
differences from standard HIRES processing at IPAC. Foremost is improved
destriping and zodiacal emission subtraction, which lead to 
reduction of artifacts, enhancement of faint structure, and
the ability to mosaic images without edge discontinuities. 
The IGA is well suited to high-resolution studies of extended structure,
and will be valuable for  a wide range of
scientific studies, including:
the structure and dynamics of the interstellar medium (ISM);
cloud core surveys within giant molecular clouds;
detailed studies of HII regions and star forming regions;
determination of initial mass functions (IMFs) of massive stars;
and study of supernova remnants (SNRs).
The IGA will be especially useful for multi-wavelength studies
using the many Galactic plane surveys that have similar ($\sim 1'$) resolution. 
These include the new FCRAO CO(1-0) spectral line (\cite {Heyer96a})
and DRAO HI line / 21 cm continuum surveys (\cite {Normandeau97}).

All image reconstruction algorithms have their quirks.
This paper describes and characterizes the IGA so
that it will be useful for quantitative scientific study.
Section \ref{desc-atlas} describes the geometry and information content
of the atlas images. Section \ref{desc-proc} gives a description on
the various processing stages, namely the basic algorithm, subtraction
of zodiacal emission, and coordinate transform and reprojection.
Section \ref{char} discusses the characteristics of the images, including
resolution, photometric and positional accuracy, mosaic properties, and
calibration. Section \ref{artifacts} details the various image artifacts.

The IGA images are available upon request from IPAC (info@ipac.caltech.edu; 
http://www.ipac.caltech.edu) or through the NASA National Space Science
Data Center (NSSDC; http://nssdc.gsfc.nasa.gov/nssdc/nssdc\_home.html).
The complete archive is comprised of ten 8mm tapes. Casual users should direct
requests for specific images to IPAC.
This paper gives references to online resources
(mostly in the form of World Wide Web documents) whenever appropriate. The 
information is accurate as of 1997.

\section{Description of the Atlas}\label{desc-atlas}

The atlas consists of images (1st and 20th iteration)
and ancillary maps in FITS (\cite{FITS81})
format in the 60 and 100 {\micron} wavelength bands. The Galactic plane 
images cover $0^\circ \leq l < 360^\circ$ in Galactic longitude and
 $-4.7^\circ < b < 4.7^\circ$ in latitude.
The field of view for each image is $1.4^\circ \times 1.4^\circ$, 
on $1^\circ$ centers
in both the Galactic longitude and latitude directions, the pixel size is
15{\arcsec}, and Galactic coordinates and Cartesian projection are used.
The molecular cloud fields (Orion, {\roph}, Taurus-Auriga) are 
rectangular, about $20^\circ$ on a side, with boundaries
selected to encompass the diffuse infrared emission of each cloud.
The images are $2.5^\circ \times 2.5^\circ$ on $2^\circ$ centers
with 15{\arcsec} pixels and  
use Equatorial coordinates (B1950) and Cartesian projection.
The boundaries are given by: 
Orion [$5^h 8^m , 6^h 12^m$ ] [-13\fdg 0,18\fdg 0], 
{\roph} [$15^h 31^m ,17^h 00^m$], [-33\fdg 0,-17\fdg 0], and 
Taurus-Auriga [$3^h 48^m ,5^h 12^m$], [12\fdg 0,33\fdg 0].
References to IGA sources in the Galactic plane should follow the sample
format IGA~G$218.50-0.50$, while the names of sources in molecular clouds
should follow the {\IRAS} format, e.g. IGA B04302+2538 to represent the position
($4^h 30\fm 2, 25^\circ 38'$, B1950).
The Cartesian projection used with Equatorial coordinates is a
relatively new FITS combination which may cause trouble for some FITS reading
software. For individual images or small mosaicked fields the FITS header 
keyword {\tt -SIN} can be used in place of {\tt -CAR} (see Sec. \ref{reproj}).

The 1st iteration images are coadded {\IRAS} images (i.e. FRESCO images)
with no resolution enhancement. They have the native {\IRAS} resolution 
of approximately $2.0' \times 4.7'$ at 60 {\micron}
and $3.8' \times 5.4'$ at 100 {\micron}.
After MCM processing to 20 iterations the typical
spatial resolution improves to $1.0' \times 1.7'$ at 60 {\micron}
and $1.7' \times 2.2'$ at 100 {\micron} (see Sec. \ref{res}).
The images at 60 and 100 {\micron} have inherently different 
resolutions.  Ratio or color maps should only be attempted by expert users,
and only after correcting the images to a common resolution.

Aperture photometry is accurate to about 25\% (see Sec. \ref{abscal}). Most 
of the uncertainty is due to background measurement uncertainties.
The images are on the same absolute flux level as the ISSA
images, except for a constant AC/DC factor 
(see Sec. \ref{acdc}).

The ancillary maps include the correction factor variance (CFV) map,
the photometric noise (PHN) map, coverage (CVG) map, the
detector track (DET) map, and the beam sample map (BEM).
See Table \ref{tmaps}
in Section \ref{MCM} for the quantities they represent and
Section \ref{example} for example images.
The FWHM.txt text file gives Gaussian beam sizes derived
from the corresponding BEM map.

\section{Description of Processing}\label{desc-proc}

For an overview of the HIRES processing developed at IPAC, see
http://www.ipac.caltech.edu/ipac/iras/hires\_over.html. This section
emphasizes the unique problems encountered in the IGA
production.

\subsection{Overview of the Production Pipeline}\label{pipeline}


{\IRAS} detector data, known as CRDD (Calibrated, Reconstructed Detector
Data), grouped in {\sevbysev} plates, reside in the
{\em ``Level 1 Archive''} at IPAC. The first step in the pipeline for mass
production of HIRES images is to extract data covering a specific field 
from the archive and then perform calibration and various other preprocessing.
We take the {\sevbysev} preprocessed and calibrated 
plates and use the algorithm described
in Section \ref{dezody} to subtract the zodiacal background emission.
This step requires the corresponding ISSA image as supplemental input
(SmLAUN in Fig. \ref{fig-pipe}, Section \ref{dezody}).

Following the calibration and zodiacal subtraction, the detector files
are broken into {$1.4^{\circ}\times 1.4^{\circ}$} fields,
and reprojected into Galactic coordinates (from equatorial) if required,
with field centers separated by 1 degree
(BrkDet in Fig. \ref{fig-pipe}, Section \ref{reproj}).
The factor-of-two overlap is a conservative insurance
against discontinuity across field boundaries (see Section \ref{mosaic}), 
as local destriping and
different flux bias (see Sec. \ref{fbias}) will be applied to each small field.
The {$1.4^{\circ}\times 1.4^{\circ}$} size is also the maximal field size with
complete coverage allowed within one Level 1 plate, given the 2 degree
redundancy of the plates and arbitrary location and orientation of the 
small field relative to the Level 1 plate. Figure \ref{fig-geom} illustrates
the overlapping IGA fields, and the geometry and orientation of the
Level 1 plates that determine the allowed IGA field size.


All operations described above are carried out on workstations.
The small field ({$1.4^{\circ}\times 1.4^{\circ}$})
detector files are then processed into HIRES
images, which is done on the Intel Paragon supercomputer.
The basic algorithm for image reconstruction
is described in Section \ref{MCM}. For parallelization
strategy and details of the destriping algorithm, see Paper~I.

\subsection{The Maximum Correlation Method}\label{MCM}

Starting from a model of the sky flux
distribution, the HIRES MCM algorithm folds the model through the
{\IRAS} detector responses, compares the result track-by-track
\footnote{Track, also called {\em leg} or {\em scanline}, refers
to the set of data samples collected consecutively by one detector
moving across a given field.}
to the observed flux, and calculates corrections to the model.
One important characteristic is that the standard MCM algorithm
conserves flux. We give a brief description of the MCM algorithm
following the formalism and notations of Aumann, Fowler, and Melnyk (1990)
\markcite {MCM90}.

Given an image grid $f_j$, with $n$ pixels $j=1,...,n$ and $m$
detector samples ({\em footprints}) with fluxes
$D_i: i=1,...,m,$
whose centers are contained in the image grid, an image can be
constructed iteratively from a zeroth estimate of the image,
$f_j^0={\rm const.} > 0$ for all $j$. In other words the initial
guess is a uniform, flat, and positive definite map. For each
footprint, a correction factor $C_i$ is computed as,
\begin{equation}
C_i = D_i / F_i,
\end{equation}
\noindent where
\begin{equation}
F_i = \sum_j r_{ij} f_j,
\end{equation}
\noindent and $r_{ij}$ is the value of the $i$th footprint's response function
at image pixels $f_j$. Therefore $F_i$ is the current estimate of
the $i$th footprint's flux, given image grid $f_j$.

A mean correction factor for the $j$th image pixel is computed
by projecting the correction factor for the footprints into the image domain:
\begin{equation}\label{c_proj}
c_j = [\sum_i (r_{ij}/{\sigma_i^2}) C_i] / [\sum_i (r_{ij}/{\sigma_i^2})].
\end{equation}
\noindent The weight attached to the $i$th correction factor for the $j$th
pixel is $r_{ij}/{\sigma_i^2}$, where $\sigma_i$ is the {\it a priori}
noise assigned to the $i$th footprint.

The $k$th estimate of the image is computed by
\begin{equation}
f_j^{(k)} = f_j^{(k-1)} c_j.
\end{equation}
\noindent In practice when the footprint noise $\sigma_i$ is not easily
estimated, an equal noise value for all footprints is assumed,
and the MCM is identical to the Richardson-Lucy algorithm
(\cite{Rich72,Lucy74}).


Table \ref{tmaps} shows the quantities represented by the ancillary maps
(\cite{MCM90}).
For more detailed information on the HIRES ancillary maps, see
http://www.ipac.caltech.edu/ipac/iras/hires\_maps.html.
The correction factor variance (CFV) 
map gives an estimate of the level of convergence at a certain
pixel, measuring the agreement of correction factors projected onto it
from different detector footprints. The photometric noise (PHN)
 map signifies the photometric
noise at a pixel, propagated from noise in the detector measurements.
The coverage (CVG) map is the sum of the response function grids of all 
footprints
within the field. The detector track (DET) map registers the footprint 
centers and helps visualize the detector scanning pattern.
Artifacts due to low coverage may be diagnosed using the
coverage maps. The remaining ancillary maps provide diagnostics
for other less frequent artifacts.

The effective beam size in HIRES images depends on the response
function and sample density in a complicated fashion, and may vary
by factors of three over distances of several arcminutes (\cite{Fowler94}).
In order to estimate the beam size at any given position and to see typical
variation over the field, ``beam sample maps'' (BEM) are provided. These are
produced from simulated detector data based on actual coverage geometry,
with the simulation scene being a collection of spike sources against
a smooth background. An image of the reconstructed spikes (beam sample
map) is generated with all the same processing options as the actual image.

\subsection{Subtraction of Zodiacal Emission}\label{dezody}

Zodiacal dust emission is a prominent source of diffuse emission
in the {\IRAS} survey. The zodiacal contribution to the observed surface
brightness depends on the amount of interplanetary dust along
the particular line-of-sight, an amount which varies with the
Earth's position within the dust cloud. Consequently, the sky
brightness of a particular location on the sky as observed by {\IRAS},
changes with time as the Earth moves along its orbit around the Sun.
The different zodiacal emission level in different scanlines, if not
subtracted, can cause step discontinuities in the images if adjacent
patches of sky were observed at different times.

A physical model of the zodiacal foreground emission based on the
radiative properties and spatial distribution of the zodiacal dust
was developed by Good (1994)\markcite{Good94}.
The {\IRAS} Sky Survey Atlas (ISSA; \cite {ISSA94}) made use of this
model and subtracted the predicted
zodiacal emission from the detector data before co-adding them.

However, the {\IRAS} detector data which serve as input to the IGA and other 
{\IRAS} image products, still contain zodiacal emission.
A preprocessing method has been developed to bring the raw detector data flux
to a common level with the ISSA images, effectively subtracting
the zodiacal emission component (Paper~I).
Nearby ISSA images ($12.5^\circ \times 12.5^\circ$, 
1.5{\arcmin} pixels) were reprojected and mosaicked 
to cover the same field-of-view as a Level 1 plate ({\sevbysev}, 1{\arcmin}
pixels). A set of simulated data is then calculated from the mosaicked
image, by running the actual {\IRAS} scan pattern through this image,
\begin{equation}
F_i^{\rm ISSA} = \sum_j r_{ij} f_j^{\rm ISSA}
\end{equation}
The difference between these simulated data and the real data is then 
used to determine the local zodiacal emission
\begin{equation}\label{eq-D_ZODY}
D_i^{\rm ZODY} = {\rm median} (D_i - F_i^{\rm ISSA})
\end{equation}
where the median is taken for nearby footprints in the same scanline
with a total spatial range of 1{\deg}.
The zodiacal component is then subtracted
\begin{equation}
D_i^{\rm NEW} = D_i - D_i^{\rm ZODY}
\end{equation}
and the new data output for use in image construction.

Because of the large spatial scale used in Eq. (\ref{eq-D_ZODY}),
the resulting zodiacal emission flux, $D_i^{\rm ZODY}$, varies smoothly
with a characteristic scale of $\sim$ 1{\deg}. Therefore the
zodiacal subtraction process does not interfere with the high
spatial frequency information inherent in the raw data, which is
needed for the image reconstruction and resolution enhancement.

\subsection{Coordinate Transform and Reprojection}\label{reproj}


Each Level 1 plate covers a field of view of {\sevbysev}, using
a projection center local to the plate.
The positions of detector footprints are stored in equatorial coordinates 
using Cartesian projection
(\cite{GW96}; FITS keywords {\tt RA---CAR, DEC--CAR}, B1950):
\begin{eqnarray}
x &=& \phi, \cr
y &=& \theta,
\end{eqnarray}
where $\theta$ and $\phi$ are angles in the native coordinate
system (Euler angles with respect to local great circles).
Each Level 1 plate has its own projection center 
($C$ in Fig.~\ref{fig-reproj}).

For the IGA, the Cartesian projection (FITS keywords
{\tt GLON-CAR} and {\tt GLAT-CAR})
with reference point at the Galactic center is convenient, 
in which case $l$ and $b$ map linearly to $x$ and $y$.

To transform the equatorial coordinates of footprints stored in the
Level 1 archive to Galactic, the following steps are done in BrkDet.
For each footprint centered at $P$, a unit vector $OP$ is computed in the 
equatorial system,
using RA and Dec of the projection center $C$, and the $x$ and $y$
of $P$ in the Cartesian projection system centered at $C$. 
Then the unit vector is rotated to the Galactic
system, and $l$ and $b$ are obtained (see Fig. \ref{fig-reproj}).
Coordinates and fluxes of footprints falling in 
each $1.4^\circ \times 1.4^\circ$ field of view are grouped together 
and written out for the final image reconstruction step.

The tilt angle for each scanline, which is necessary for calculating the
response function grid during image reconstruction, also needs to be
redetermined in the Galactic coordinate system. For a scanline with
$n$ footprints located at $(x_i, y_i), i = 1,...,n$, 
this was done by fitting
a straight line through the $x$ and $y$ values, by minimizing 
$\sum_i \Delta_i^2$, where $\Delta_i$ is the distance from footprint $i$
to the line. This gives the estimate for the tilt angle $\Phi$, measured
relative to the $x$ axis
\begin{equation}
\Phi = {1 \over 2} \arctan { 
                             {2 \sum_i x_i y_i - {\overline x}\,{\overline y} }
  \over
                             {\sum_i x_i^2 - {\overline x}^2 
                                   - y_i^2 + {\overline y}^2 }
                           }
\end{equation}
where 
\begin{eqnarray}
{\overline x} &=& \sum_i x_i / n \cr 
{\overline y} &=& \sum_i y_i / n.
\end{eqnarray}

For the molecular cloud fields (Orion, {\roph}, and Taurus-Auriga),
equatorial coordinates were used (FITS keywords {\tt RA---CAR} and
{\tt DEC--CAR}, B1950), and the Level 1 archive geometry was retained
(no reprojection of the footprint data was performed).
Each Level 1 plate ({\sevbysev} on $5^\circ$ centers) was divided 
into 3 $\times$ 3 subfields of
$2.5^\circ \times 2.5^\circ$ each, on $2^\circ$ centers, with the
projection center the same as the Level 1 plate center.
Therefore the subfield images 
belonging to the same Level 1 plate are mosaickable
without the need of reprojection, but special care needs to be taken
when mosaicking subfields from different Level 1 plates.
The use of Cartesian projection for Equatorial coordinates
is closest to the native format of the {\IRAS} data, but differs
from the more commonly used  {\tt -SIN} projection by about 0.5 pixel 
at the edge of a Level 1 plate.

\subsection{Issues Related to Flux Bias}\label{fbias}

Astronomical images often contain backgrounds which need to be subtracted from
the source of interest, such as instrumental offsets, sky backgrounds, or in
the case of the {\IRAS} data, zodiacal light or diffuse Galactic backgrounds.
This means that the zero level of an image depends on the specific application 
for which part of the signal is considered source or background. Taking 
advantage of this arbitrary nature of the background level, many resolution 
enhancement schemes add a constant value to the image which is chosen to optimize
the performance of the algorithm.

Because of the nonlinear nature of the MCM algorithm, the spatial resolution
achieved by HIRES processing is not invariant under application of
an additive {\em flux bias}:
\begin{equation}
D_i \longrightarrow D_i + F_{\rm BIAS}.
\end{equation}
Generally, the closer to zero the data, the higher the resolution obtained.
Alternatively, the more iterations, the higher the resolution obtained.
However, the MCM algorithm is unstable to negative data values. The optimum
performance is obtained by using data with small but nonnegative values.
Therefore, to maximize both spatial resolution and throughput, 
a flux bias is computed and applied before
the image reconstruction step, to bring the data close to zero in order 
to achieve higher resolution at a given iteration. Since the flux bias is
only important during the image processing step,
the applied flux bias is subtracted from the result image
so that the surface brightness of the output image matches the original
data.

For IGA processing, the flux bias is calculated in the BrkDet step,
using the negative of the first percentile from the flux histogram
in each $1.4^\circ \times 1.4^\circ$ field.
In other words, the first percentile is used as the zero point in
subsequent HIRES processing. The detector data having fluxes below
the first percentile are discarded, since negative data 
cause instabilities in the algorithm.

\section{Characteristics of the Images}\label{char}

In this section ``IGA(1)'' denotes the 1st iteration IGA images,
and ``IGA(20)'' the 20th. The resolution, photometric accuracy,
positional accuracy, surface brightness accuracy, mosaic property,
and residual hysteresis effect of the images are discussed and
quantified.

\subsection{Resolution}\label{res}

The diffraction limit of the {\IRAS} 0.6m telescope is 50$''$ and 84$''$ at
60 and 100 microns, respectively. The effective beam of
the coadded {\IRAS} data is much larger, typically 
$2.0' \times 4.7'$ at 60 {\micron}
and $3.8' \times 5.4'$ at 100 {\micron}
due to the large and rectangular {\IRAS} detectors.
The MCM algorithm makes use of the geometric information in
the large number of redundant tracks with differing scan angles
to extract higher spatial resolution, which in some cases
can approach the diffraction limit of the telescope (\cite {Rice93}). 
The effective beam size in HIRES images depends on the response
function and sample density in a complicated fashion.
The resolution also depends on the magnitude of the point source
relative to the effective background (see also Sec. \ref{fbias}).

The spatial resolution of a given field can be estimated from the
corresponding BEM maps produced from simulated data
(see also Sec. \ref{MCM} and Table \ref{tmaps}). To generate the
BEM maps artificial point sources are added to the smoothed data, which 
then undergo HIRES processing. Specifically, point sources are identified
and removed from IGA(20) image: the image, further smoothed, provides 
a model background to which regularly spaced (12$'$) point sources are added.
The magnitude of the planted point source spikes is adjusted according to the
dynamic range of the IGA(20) image: the pixel intensity 
is set to $ 20 \times (99\% {\rm quantile} - 50\% {\rm quantile}) $
of the IGA image histogram (plus the background). The numerical value of 20
approximately converts the flux from per unit beam to flux per unit pixel.
This arbitrary choice of flux is meant to represent a typical point source 
which is strong enough with respect to the local Galactic background to 
benefit from high-resolution processing.
A set of simulated data is then generated from the artificial image,
from which the BEM map is reconstructed through HIRES.
A Gaussian profile is fitted to the reconstructed point sources in
the BEM maps,
and the FWHM along the major and minor axes are taken as the
measure for the achieved resolution.


The typical resolution of the IGA(20) images is 
$1.0' \times 1.7'$ at 60 {\micron}
and $1.7' \times 2.2'$ at 100 {\micron}, which represents a
substantial improvement over the coadded images.
Figure \ref{fig-res1} demonstrates the dependence of the resolution upon
longitude across the Galactic plane. The plotted major and minor axis
FWHM were averaged over latitude. Two obvious dips are seen in the
major axis curves, both in 60 and 100 {\micron}, near $l = 100^\circ$ and
$l = 280^\circ$. These two areas in the Galactic plane featured 
near-perpendicular intersecting
scanlines in the {\IRAS} survey, and the extra geometric information in
the data gives rise to the increased resolution.

To investigate the dependence of resolution on source strength relative 
to background level, BEM maps were generated for simulated point sources 
ranging from 1 to 10,000 Jy in strength. The background intensity level 
determined for the test field near $l =$ 120{\deg} 
was 53.94 MJy/sr and 165.05 MJy/sr at 60 and 100 {\micron}, respectively. 
Integrated over the effective solid angle of the {\IRAS} detectors,
$6.25 \times 10^{-7}$ and $13.54 \times 10^{-7}$ sr,
the detector fluxes due to the local background become 33.7 and 223 Jy, 
respectively.
To find the effective background during HIRES processing,
the flux bias value from the FITS header (see Sec. \ref{fbias})
can be converted from W m$^{-2}$ to Jy through division by the 
conversion factors $2.58 \times 10^{-14}$ or $1.00 \times 10^{-14}$
at 60 and 100 {\micron}, respectively, and then subtracted from the 
corresponding local background 
flux. In this test case, zero flux bias was used, giving simply
33.7 and 223 Jy for the processing background at 60 and 100 {\micron}, 
respectively. We emphasize that the spatial resolution depends on
the effective background level {\it during} HIRES processing, namely the
local background minus the flux bias value. A quick way to estimate 
the processing background in IGA images is to find the minimum intensity 
value in the image (which will be close to the flux bias value) and then 
subtract it from the local background. If necessary, convert from intensity to
flux in Jy using the detector solid angles given above.

The results plotted
in Figure \ref{fig-res2} show that the IGA(20) resolution 
is at least a factor of two better than the coadded IGA(1) 
resolution. Also, the resolution significantly improves
for point sources stronger than the processing background
of 33.7 and 223 Jy at 60 and 100 {\micron}, respectively. 
Furthermore, when the
source-to-background contrast reaches about 20, the achieved resolution 
becomes insensitive to the background. The resolution in other 
fields/regions should behave in
the same qualitative fashion when the local processing background 
is computed as above.

Figure \ref{fig-res2} shows the additional effect that 
offset compensation destriping (Paper~I, also 
see Sec. \ref{stripe}) gives comparable but slightly poorer resolution
than standard HIRES destriping,
especially along the major axis (cross scan) 
direction. 

The number of iterations also effects the spatial resolution, although in 
practice most of the improvement in spatial resolution is gained within the 
first 10 iterations.
For IGA processing, 20 iterations was chosen in a tradeoff between
speed of processing and artifact development versus spatial resolution.
However, for strong sources, regions of high coverage, or regions with
favorable scanline geometry, data reprocessing with additional
iterations gives better spatial resolution
(\cite {Rice93,Hurt96}).


In addition, note that the beams of the IGA images are not
Gaussian. The most prominent deviation of the beam from a 2-dimensional
Gaussian is due to the ringing artifact (Section~\ref{ring}). Rice (1993)
gives a detailed account of the HIRES beams.

\subsection{Photometric Accuracy}\label{photo}
To test the photometric accuracy
thirty-five relatively isolated point sources (with a well-defined background)
were selected. All sources have flux $>$ 10 Jy and are spatially unresolved 
as measured by the
Correlation Coefficient (CC) flag in the {\IRAS} Point Source Catalog (1988).
Fluxes were measured using an aperture photometry
program developed at IPAC, in which
the median pixel intensity within an annulus (radius 5{\arcmin}--7{\arcmin})
centered at the point source position (taken from the PSC)
is taken as the background intensity. Two estimates of the point source
flux are then made, using the total fluxes within 5{\arcmin} and 7{\arcmin}
radius from the PSC position (minus the background intensity $\times$
the number of pixels).
For sources chosen for the photometry test,
these two values are usually sufficiently close to each other to indicate
a well-defined background level.
The average of these two values is taken as the point source flux from
the IGA image, and compared against the value from the PSC.
The computed fluxes are given in Table \ref{tphoto}, while 
Table \ref{tphotostat} summarizes
the statistical correlation between IGA and PSC flux values.

An overall offset (12$\%$) between the IGA(1) and PSC fluxes is seen
at 60 {\micron} which is, however, not present at 100 {\micron} (1$\%$).
One possible explanation for the 60 {\micron} offset is the
different data calibration used, specifically the hysteresis
correction. The IGA and other recent {\IRAS} image products are
based on the final {\IRAS} Pass 3 calibration, described in detail
in the ISSA Explanatory Supplement (\cite{ISSA94}). This calibration includes
a hysteresis correction at both 60 and 100 {\micron} (see Sec. \ref{hyst}).
The PSC, however, is based on {\IRAS} Pass 2 CRDD data, which were corrected
after the fact to the Pass 3 calibration. One significant difference
is the way in which hysteresis was treated:
the PSC applied a hysteresis correction only at 100 {\micron} 
(\cite{IRAS88}). 
The lack of hysteresis correction at 60 {\micron} for PSC sources,
particularly important for the Galactic plane where hysteresis is strongest, 
provides one explanation for why 
there is a systematic offset at 60 {\micron}, but not at 
100 {\micron}, between the IGA(1) and PSC fluxes.

In addition, Table \ref{tphotostat} shows there is a growth 
in flux from the 1st to 20th iteration which is small
(2$\%$) for the 60 {\micron} band, but significant (10$\%$) 
for the 100 {\micron} band. Analysis shows that the effect is caused by 
the depression of the background due to the ringing artifact. 
In the Galactic plane, where the background emission
is strong and structured, the largest contributor to the flux uncertainty 
is the background determination (\cite{Fich96}).
The total flux within the selected aperture is comprised of the source
flux plus a background contribution (background $\times$ area). For
the photometry sample, the ratio of background flux to source flux is
1.8 at 60 {\micron} and 6.7 in the 100 {\micron} band.
In addition, the background level systematically decreases on average
by 1.7$\%$ for 60 {\micron}, and 1.8$\%$ at 100 {\micron} due
to ringing in the 5$'$ -- 7$'$ annulus. This leads to an apparent flux increase 
from 1 to 20 iterations 
of $1.8 \times 1.7\% = 3\%$ at 60 {\micron} and $6.7 \times 1.8\% = 12\%$ in
the 100 {\micron} band, which agrees with the results of 
Table \ref{tphotostat}.

To compensate for the systematically low background levels we
recomputed the IGA(20) source fluxes using IGA(1) background levels.
The resulting fluxes show no systematic offset (mean of IGA(20)/IGA(1) = 0.99)
and better correlation with IGA(1) fluxes (standard deviation
= 0.10). This technique of using IGA(1) backgrounds to calculate IGA(20)
fluxes is recommended whenever the most stable and accurate photometry 
is required.


The growth in flux found for IGA point sources is not a universal
property of HIRES processing. In a HIRES study of interacting Galaxy
pairs, Surace et al. (1993) found HIRES fluxes systematically decreased
by 20\% from iteration~=~1 (FRESCO) to iteration~=~20, a result they
attributed to the small extended nature of the sample.
Since the MCM algorithm fundamentally conserves flux, the effect
is either due to a systematic increase in the background, or
to redistribution of flux outside the photometric aperture. The use
of the IGA(1) background to compute the IGA(20) flux is a technique that
can help determine the cause of such systematic trends.

Figure \ref{fig-photo} plots 
the dependence of (IGA(20) flux / PSC flux) on the PSC flux.
There is no trend with source flux, apart from the previously discussed 
offsets. 

\subsection{Size Dependent Flux Correction}\label{acdc}

The estimation of the flux for extended sources ($> 4'$ -- $40'$) may involve
a size dependent flux correction, also known as the AC/DC correction.
The {\IRAS} detectors had a dwell-time dependent responsivity
change. Hence, the gain changes as a function of source size: at the
{\IRAS} survey speed of 3.85 arcmin/s, the gains leveled off for
structure on the order of 40{\arcmin} in extent. This effect was
band-dependent and largest at 12 {\micron}. Thus, there are
two calibrations for the {\IRAS} data, the calibration appropriate for
point sources, known as the AC calibration, and the calibration appropriate
to very extended structure, known as the DC calibration. To
bring point source fluxes measured from DC-calibrated products to the AC
(same as the PSC) calibration, the fluxes
must be divided by 0.78, 0.82, 0.92 and 1.0 at 12, 25, 60 and
100 {\micron}, respectively. The IGA uses the AC calibration,
while the ISSA images are on the DC scale.

Point source fluxes obtained by aperture photometry with
appropriate background subtraction on AC calibrated images such as the IGA
should be consistent
with the PSC. However, neither calibration is strictly correct 
for structure on spatial scales intermediate between point sources 
and 30{\arcmin} in size. Intermediate-scale corrections and
uncertainties can be estimated from the
plots in the {\IRAS} Catalogs and Atlases: Explanatory Supplement (1988).

\subsection{Calibration Uncertainty}\label{abscal}

The flux measurement uncertainty derived from point sources 
(Table \ref{tphotostat}) for the
IGA(1) images is 11\% and 16\% at 60 and 100 {\micron},
respectively. Given the
isolated nature of the sources, these uncertainties represent
a best case. A better estimate of the measurement uncertainty
in more complex regions is given by
Fich \& Terebey (1996), who find 17\% and 18\% for the flux measurement 
uncertainty of at 60 and 100 {\micron}, respectively, for a sample
of outer Galaxy star forming regions.

In some cases, systematic instrumental effects also contribute 
significantly to the flux uncertainty. The {\IRAS} calibration
for point sources is accurate, albeit affected by residual hysteresis
at 60 and 100 {\micron} in the Galactic plane. As described in
the previous section, there is a 12\% systematic uncertainty
at 60 {\micron} between the IGA(1) and PSC. At 100 {\micron} the
uncertainty due to residual hysteresis is less than 5\% over most
of the Galactic plane, but approaches a maximum of 20\% near 
the Galactic center.

For small but extended ($5'$ -- $20'$) sources the situation is complex.
The size-dependent flux correction, the so-called AC/DC effect
(see Sec. \ref{acdc}) is typically about 10\% or less.
However the detector response is not well-behaved for bright
extended sources: above 100~Jy the {\IRAS} Explanatory Supplement 
quotes uncertainties of 30\% at 60 {\micron} and 70\% at 100{ \micron}
(\cite{IRAS88}).

Prominent in the IGA is diffuse Galactic emission associated with
HI, which varies on a scale of a few degrees. The {\IRAS} -- COBE 
comparison gives an indication of the calibration uncertainty.
Over angular scales larger than 10{\deg} the {\IRAS} calibration differs
systematically from that of COBE by 13\% and 28\% at 60 and 100 {\micron},
respectively (\cite{ISSA94}).

The ISSA survey was used as large scale surface brightness truth table
for the IGA. This implies that defects or uncertainties introduced
by the ISSA processing extend to the IGA as well (see Sec. \ref{discont}). 
At 60 and 100 {\micron}, residuals associated with zodiacal emission model 
can approach 1 -- 2 MJy/sr in the ecliptic plane (Galactic center and 
anti-center directions),
but are typically far less (e.g. \cite{Fich96}). For more information
consult the ISSA Explanatory Supplement (\cite{ISSA94}).

\subsection{Positional Accuracy}\label{pos}

{\IRAS} Point Source Catalog positions were used as truth tables for
a positional accuracy test of the IGA.
Positions were computed for the same sample of thirty-five sources used in
the photometry comparison. For each source, a circular area
with radius 5{\arcmin} was defined (centered at the PSC position),
and the area's flux-weighted centroid was taken as the point source
position implied by the IGA image and compared against the PSC position.

Table \ref{tpos} shows the result of the comparison.
For the 60 {\micron} band, the distances between the IGA position
and PSC position have an average of 7.6{\arcsec} and standard deviation
5.6{\arcsec}, while for 100 {\micron}, there is a 
7.1{\arcsec} $\pm$ 4.1{\arcsec} difference.

The PSC reports error ellipses corresponding to the 95\% confidence 
level for source positions. The major and minor axes of the error ellipse
correspond approximately to the cross- and in-scan directions. 
For each source, we projected the IGA position along 
the major and minor axes of the error ellipse centered at the PSC position.
The mean deviations from the PSC position were found to be 
similar along the major and 
minor axis directions, and do not scale with the length of the major and
minor axes. This indicates the positional errors produced by the MCM algorithm 
are due to nonsystematic effects unrelated to the {\IRAS} scan pattern and
detector geometry.

\subsection{Surface Brightness Accuracy}

To test the surface brightness of the zodiacal light subtracted IGA images,
they were rebinned to match the ISSA geometry and compared pixel-by-pixel 
against the ISSA images. The standard deviation of the 1.5{\arcmin} 
pixel-by-pixel difference is less than 6\% for IGA(1) vs. ISSA, and less 
than 12\% for IGA(20) vs. ISSA. No systematic offset was found
between the IGA and ISSA data. See Paper~I for details.


\subsection{Mosaic Property}\label{mosaic}

The geometry of the IGA images allows them to be mosaicked without
any reprojection, hence no smoothing is required and the original
resolution can be retained in the mosaicked images. To reduce  edge
discontinuities, the images should first be cropped to 
$1^\circ \times 1^\circ$ from $1.4^\circ \times 1.4^\circ$
with the centers unshifted before mosaicking. No offset needs to be
applied to the different subfields. In most cases the mosaicked image
is seamless to the human eye.

Quantitatively, within a chosen Level 1 plate in the W3, W4, and W5 region, 
pixel intensity ratios were calculated for 1-pixel wide
edges covered by neighboring subfields after cropping the subfields
 to slightly larger than $1^\circ \times 1^\circ$. 
Table \ref{tmos} summarizes the intensity ratio statistics for
both the 1st and 20th iteration images.
A total of 10122 pixels in 42 1{\deg} edges
were used in the calculation.
For 20th iteration images, the standard deviation of the ratio amounts 
to 0.51\%  and 0.23\%
for band 3 (60 {\micron}) and 4 (100 {\micron}) respectively.
Intensity ratio statistics were also calculated for cross-Level 1-plate 
boundaries, using a total of 8194 pixels in 34 1{\deg} edges.
Again for 20th iteration images, the standard deviations are
1.5\% and 0.46\% for band 3 and 4. The match is worse than
that of intra-plate edges, since the zodiacal subtraction was done
separately for each Level 1 plate (see Sec. \ref{dezody}).

The better boundary match (smaller deviation) at band 4 can be understood from
the poorer resolving power of HIRES at band 4 than at band 3, which
decreases the resolution difference between subfields caused by the
different flux bias levels used in the image reconstruction process
(see Sec. \ref{fbias}).

\subsection{Residual Hysteresis}\label{hyst}

The {\IRAS} detectors showed photon induced responsivity enhancement,
known as the hysteresis effect, especially in the 60 and 100 {\micron}
bands. The effect is prominent when the scanlines pass the Galactic plane
(e.g. \cite{IRAS88}, Chap. VI) and thus a concern for the IGA survey. 
The final {\IRAS} Pass 3 calibration, on which both the IGA and ISSA are
based, employed a physically based detector model to correct for the 
hysteresis. However, the technique could not correct variations that were 
more rapid than $\sim$ 6{\deg} in spatial scale (\cite{ISSA94}, Chap. III).
This section quantifies the residual hysteresis near the Galactic plane
in the ISSA data, which should also describe the residual
hysteresis present in the IGA.

In the {\IRAS} survey,
a given region can be covered by up to 3 scans carried out at different
times, known as Hours CONfirming (HCON) scans. HCON 1 and HCON 2 were
separated by up to several weeks, while HCON 3 was taken roughly 6 months
later. 
\footnote{Most ($96\%$) of the sky was covered by at least two
HCONs, and $2/3$ of the sky was covered by three HCONs.}
This meant HCON 3 usually passed the Galactic plane along the opposite
direction of HCON 1 and 2, since {\IRAS} followed a Sun-synchronous
orbit and the telescope always pointed approximately 
90{\deg} away from the Sun.


Figure \ref{fig-hyst-exp} illustrates the effect on computed flux values
from the different HCONs that is caused by the photon induced responsivity 
change. At the starting and ending points of a scan, {\IRAS} detectors were lit
up by an internal calibration flash, which anchored the responsivity of the
detectors at these two points. In the early calibration schemes, the response
was assumed to change linearly between the two calibration flashes, which
imperfectly tracks the true detector response
when the scan passes through bright regions like the Galactic plane
(Fig. \ref{fig-hyst-exp}a).
Figure \ref{fig-hyst-exp}b illustrates the resulting deviation of the
computed fluxes from true values. For example, the computed flux of a 1 Jy
source differs from 1 Jy in a systematic way that depends both on galactic
latitude and the scanning direction. Figure \ref{fig-hyst-exp}c shows the ratio
of fluxes determined from descending and ascending scans; the ratio should equal
one for perfectly calibrated data.

To quantify the residual hysteresis effect in the ISSA images, 
ISSA images made from HCON 1 and 3 were compared at $l =$ 0{\deg}, 
10{\deg}, 20{\deg}, 60{\deg}, 120{\deg}, 180{\deg}, 240{\deg}, 300{\deg}, 
340{\deg}, and 350{\deg}. Images covering $\pm 5${\deg} latitude and
$\pm 2.5${\deg} longitude were first smoothed with a 4.5{\arcmin} boxcar
kernel, roughly the ISSA resolution, and then summed over 5{\deg} longitude
intervals to increase signal to noise.
Pixel intensity ratios (HCON 1 / HCON 3) were computed then averaged over
each 5{\deg} ($l$)$\times$ 4.5{\arcmin} ($b$) box.

Figures \ref{fig-hyst-0-120} and \ref{fig-hyst-180-300} plot the
average intensity (left column) and HCON1/HCON3 intensity ratio (right column) versus Galactic latitude. Plots made versus Galactic latitude are sufficient
for our purpose, although strictly speaking ecliptic latitude better 
represents the {\IRAS} scanning direction. The hysteresis signature is seen
clearly near $l$ = 0{\deg} with an amplitude of about 20\% at 100 {\micron}.
As expected, the peak of the average intensity plot corresponds to 
the appearance of the hysteresis signature in the ratio plot.
Hysteresis may also be present in the $l$ = 60{\deg} and 300{\deg} graphs but
below the 5\% level. Other small ($ <$ 5\%) but systematic variations in the
ISSA ratio are likely due to destriping differences. 
Figure \ref{fig-hyst-ratio} shows the maximum and minimum 
HCON1/HCON3 intensity ratio found at each longitude. At 100 {\micron} and
within 60{\deg} of the Galactic center, residual hysteresis becomes
larger than systematic differences due to destriping and noise.




\section{Artifacts}\label{artifacts}

For general descriptions of artifacts produced by HIRES processing,
see http://www.ipac.caltech.edu/ipac/iras/hires\_artifacts.html.

\subsection{Striping Artifacts}\label{stripe}

Stripes were formerly the most prominent artifacts in HIRES images.
HIRES takes as input the {\IRAS} detector data, and if not perfectly
calibrated, would try to fit the gain differences in the detector scans
by a striped image. The striping builds up in amplitude and sharpness
along with the HIRES iterations, as the algorithm refines
the ``resolution" of the stripes.

An algorithm was developed to eliminate the striping artifacts.
The basic technique involved is to estimate gain variations in the detectors
and compensate for them within the image reconstruction process. 
Observation of the Fourier power spectrum of the resulting images shows that
the algorithm eliminates the striping signal after roughly ten 
iterations. Therefore striping artifacts have been virtually eliminated
from the IGA images. See Paper~I for details and examples.


\subsection{Ringing Artifacts}\label{ring}

``Ringing'' is a prominent artifact in the IGA images.
When a point source is superimposed on a non-zero background,
the artifact known as {\em ringing} or {\rm ripples} appears in
many image reconstruction algorithms. 
In Fourier language, the reconstruction process tries to make the image
agree with the true scene in the low spatial frequency components (data
constraint),
without access to the infinitely high spatial frequencies inherent in
the point source scene.
The magnitude of the ringing depends on the strength of the point source,
the level of the residual background intensity (after the application of
flux bias), and the detector scan pattern. For nonlinear algorithms
(such as MCM) the dependence is complicated and difficult to quantify.

The ringing artifact adds uncertainty to the level of background emission
around point sources, thus hindering photometric accuracy (see Sec.
\ref{photo}). The ringing may also
interfere with the lower intensity structures present in the background.
Numerous approaches have been developed in the
field of astronomical image reconstruction to overcome the difficulty
(\cite{Frieden78,Lucy94,Bont94}).

At the time when IGA image production started, no satisfactory algorithm
was found for the purpose of ringing suppression for the atlas
(see Paper~I) which preserves
photometric integrity and does not require extra prior knowledge
(such as the positions and fluxes of point sources) as input.
Therefore the IGA images were produced with the standard MCM algorithm
(plus gain compensation destriping), which has the advantage of proven
flux conservation. Ringing thus remains as the only major
artifact in the IGA images.

Figure~\ref{fig-ring} demonstrates the ringing artifact around several point sources. At the 1st iteration, the point sources are poorly resolved and no ringing
is seen. At the 20th iteration, low intensity rings (the shape of which is
roughly elliptical and determined primarily by the detector response functions) 
surround the point sources. Further away from each point source, a brighter
ring is usually visible.


An iterative algorithm was later developed (but too late for IGA production) which aims to maximize the relative Burg entropy between modeled and measured data (\cite{Cao97,paper3}). The algorithm was run on several test fields, and was found to suppress ringing effectively and give good photometry. A partial
convergence proof has also been found. The algorithm has been applied to
the bright infrared star $\alpha$ Ori with good results (\cite{Nor97}).
At 60 {\micron} the morphology of a 7$'$ sized bow-shock shows dramatic improvement. 
On a smaller scale the ringing is suppressed to the level where diffraction spikes surrounding the star become visible.

\subsection{Glitches}\label{glitch}
Glitches are caused by hits on individual detectors by cosmic rays
or trapped energetic particles. The IPAC utility LAUNDR passes the flux
values in each scanline through two filters, one detecting point sources
and one glitches. If the ratio of the power in the point source filter
to that in the glitch filter is greater than a certain threshold
(default is 1), the phenomenon is taken to be a point source, otherwise
a glitch.


In a few regions found by visual inspection, glitches were mistaken as 
point sources and leaked into the image reconstruction stage. In such
cases reprocessing with a higher point source to glitch power threshold in
LAUNDR sufficed to eliminate the artifact.
However, it is not guaranteed that all such artifacts have been uncovered.

Remaining glitches in the IGA are rare but fairly easy to identify.
In 1st iteration images, a glitch traces out the shape of a single
detector response function, and possesses a different profile from
that of a point source (a glitch being narrower than a point source).
At the 20th iteration, a glitch would take on a ``broken-up'' shape, showing
structures finer than the physically achievable resolution, as shown
in Figure~\ref{fig-glitch}, while a point source is usually characterized
by the ringing artifact.
These differences provide a way to distinguish between real point sources
and glitches in the images.

\subsection{Discontinuities}\label{discont}

The ISSA images employed both global and local destriping techniques,
and the local destriping left some amount of intensity discrepancy
between adjacent ISSA plates (\cite{ISSA94}).

When reprojecting and mosaicking the ISSA images to the Level 1 geometry
(against which the detector data are calibrated and zodiacal emission
removed), care was taken to adjust the cropping of neighboring ISSA
plates to minimize the discontinuity. In a small number of cases, however,
some discontinuity remained which eventually affected the final IGA
image. The discontinuity is not seen in the 1st iterations, but is
sharpened and visible in the 20th. Less than $0.5\%$ of all the 
$1.4^\circ \times 1.4^\circ$ subfields are affected by this artifact.
Figure~\ref{fig-discont} shows one instance of the discontinuity across
a subfield (60 {\micron}, 20th iteration). 
The difference in intensity is approximately 5 MJy/sr.


The different flux bias values used in different $1.4^\circ \times 1.4^\circ$ 
fields also affects the mosaicking property of nearby images, since different
resolutions are achieved in the overlap region from the two images. See 
Section~\ref{mosaic} for a detailed discussion.

\subsection{Coverage Artifacts}\label{cvg}

After the processing of the mini-survey ($-1.7${\deg} $< b <$ 1.7{\deg}),
it was found that the data processing window was too small,
causing coverage depletion, and therefore unreliable structure
near window boundaries. A border of at least 5$'$ should be cropped from
images within the mini-survey.
For the extended survey ($1.3${\deg}$ < | b | <$ 4.7{\deg}),
a larger window ($1.67^{\circ} \times 1.67^{\circ}$)
was used in BrkDet to avoid coverage depletion. 
 
The use of a flux bias (see Sec. \ref{fbias}), to bring the data 
closer to zero during processing, and thereby increase throughput, was 
necessary but led to a subtle artifact. The IGA processing subtracted
a flux bias from the data corresponding to the first percentile from the 
flux histogram. Data below the threshold were discarded. This procedure 
effectively assumes the lower $1\%$ of the data
are due to noise in the flux values, which is not always justified.
In fields which had structured backgrounds, particularly at 100 {\micron},
it was found that discarding data resulted in severe coverage depletion
at the intensity minimum of an image. All images where the coverage
fell below a value of 5 in the coverage map were reprocessed with a smaller
flux bias. However problems, such as anomalous structure near the
image intensity minimum, may remain. The ancillary CVG map
can help diagnose problems associated with inadequate coverage.

The HIRES algorithm can cause systematic positional shifts if the coverage
changes abruptly. In cases where positional accuracy is important,
the CVG maps should be checked for the presence of discontinuities or
steep ($< 5'$) coverage gradients. The sense of the artifact is to
shift source positions systematically down and along the coverage gradient.

\section{Example Images}\label{example}

To illustrate the image quality of the IGA,
mosaics at 60 {\micron} of a restricted latitude range 
($-1.7${\deg} $< b <$ 1.7{\deg}) were made for
regions between Galactic longitude 280{\deg} and 80{\deg} (approximately
$16 \%$ of the total area covered by the atlas), and are shown in
Fig.~\ref{fig-0-40}, \ref{fig-40-80}, \ref{fig-280-320}, and
\ref{fig-320-0}.
Most of the emission is from stellar heated dust and shows a wealth of
star-forming regions, HII regions, and diffuse infrared cirrus 
(e.g. \cite{Fich96}).
Extended Galactic infrared emission, long associated with the Galactic HI layer,
is readily apparent as enhanced emission near the midplane 
(e.g. \cite{Terebey86}, \cite{Sodroski89}).
Each panel covers 11.5{\deg} in longitude. The dynamic range is much
larger than can be displayed, therefore
the stretch is logarithmic, with the range chosen separately for each panel
to emphasize the most structure. 






The complete set of available images and ancillary maps is illustrated
for an individual $1.4^\circ \times 1.4^\circ$ field near IC 1805 in
the second Galactic quadrant. Figure \ref{fig-ica} shows the coadded and
resolution enhanced images plus beam sample maps. Figure \ref{fig-icb}
shows the associated diagnostic ancillary maps (see Sec. \ref{desc-atlas}).
The source IC 1805, an OB cluster exhibiting strong winds and ionizing 
radiation, is located near the brightest FIR emission. To the north,
a cloud suffering erosion from
the IC 1805 cluster appears in the infrared as a cometary shaped arc 
(\cite{Heyer96b}). An HI survey of the region shows that 
the OB cluster appears to fuel a Galactic chimney (\cite{Normandeau96}).

\section{Summary}\label{summary}

The {\IRAS} Galaxy Atlas, an atlas of the Galactic plane
($-4.7${\deg} $< b <$ 4.7{\deg}) plus the molecular clouds
in Orion, {\roph}, and Taurus-Auriga, has been produced at 60 and 100 {\micron}
from {\IRAS} data. The HIRES processor, which incorporates the MCM
resolution enhancement algorithm, was ported to the Caltech parallel
supercomputers for the CPU intensive task.

At 60 {\micron} the typical resolution is 
$2.0' \times 4.7'$ for coadded IGA(1) (iteration~=~1) images, and
$1.0' \times 1.7'$ for resolution enhanced IGA(20) images,
which compares favorably with the 50$''$ diffraction limit of the {\IRAS}
telescope and the $5'$ resolution of the previously released {\IRAS} Sky 
Survey Atlas (ISSA). At 100 {\micron}, where the diffraction limit is
$84''$,  the typical IGA(1) resolution is 
$3.8' \times 5.4'$ and IGA(20) resolution is
$1.7' \times 2.2'$, again compared with the $5'$ ISSA resolution.

The IGA contains images, beam sample maps to assess 
local resolution, and ancillary diagnostic maps in FITS format. Field sizes are
$1.4^\circ \times 1.4^\circ$ in the Galactic plane, and 
$2.5^\circ \times 2.5^\circ$ in the Orion, {\roph}, and Taurus-Auriga
molecular clouds.

Zodiacal emission has been removed from the images. The result is images 
which are easily mosaicked by simple cropping and contain negligible seams.
Stripes in the images, long the limiting artifact of standard HIRES processing,
have been eliminated by algorithmic improvements to the destriping procedure.
``Ringing'' around point sources is the major artifact remaining in the
IGA images.

Photometry on the IGA images is accurate to roughly 25\%, depending on
the wavelength and size scale, while positions
agree with the {\IRAS} Point Source Catalog to better than 8$''$ standard 
of deviation.

The IGA, combined with other Galactic plane surveys 
of similar ($\sim 1'$) resolution, provides a powerful venue for 
multi-wavelength studies of the interstellar medium, star formation and
large scale structure in our Galaxy.

\acknowledgments

We are indebted to Ron Beck and Diane Engler who carried out the production
and recurring rounds of reprocessing of the IGA. 
We thank John Fowler for his help with the YORIC software.
The project received support from the Astrophysics Data Program of the
National Aeronautics and Space Administration under contract No. NAS5-32642.
This work was performed in
part at the Jet Propulsion Laboratory, California Institute
of Technology, under a contract with the National Aeronautics
and Space Administration.
The atlas production was performed in part using the Intel Paragon
operated by Caltech on behalf of the Concurrent Supercomputing Consortium.

\bibliography{refs}

\clearpage

\begin{figure}
\bigskip
\caption{Outline of the IGA Production Pipeline}
\label{fig-pipe}
\end{figure}

\begin{figure}
\bigskip
\caption{Geometry of IGA fields in Galactic coordinates relative to
the input Level 1 Plate data in equatorial coordinates. The Atlas
covers $-4.7^{\circ} < b < 4.7^{\circ}$. The small shaded areas represent
IGA fields ($1.4^{\circ} \times 1.4^{\circ}$ on 1{\deg} centers),
while the large shaded areas show Level 1 plates ({\sevbysev} on 5{\deg} 
centers). The configuration
shows an extreme case where $1.4^{\circ} \times 1.4^{\circ}$ is the largest
IGA field that can be fully covered by any single Level 1 plate.}
\label{fig-geom}
\end{figure}

\begin{figure}
\bigskip
\caption{Reprojection of Footprint Coordinates. In the diagram on the left,
RA and Dec are known for the Level 1 plate center $C$, and $x$
(negative as shown here) and
$y$ are known for the footprint $P$.
The components of the unit vector $OP$ is then computed (in equatorial
system). As shown on the right, the vector $OP$ is rotated to Galactic system,
from which $l$ and $b$ of the footprint are obtained.}
\label{fig-reproj}
\end{figure}

\begin{figure}
\bigskip
\caption{
Dependence of Beam Size on Galactic Longitude. Left and right
plots are for (a) 60 and (b) 100 {\micron} respectively. The top and bottom
curves in each figure are the FWHM of the Gaussian fitted beam along the major
and minor axes respectively. The regions in the
Galactic plane which had intersecting
scanlines in the {\IRAS} survey
are seen as two dips in the major axis curves (better resolution due
to extra geometrical information).
\label{fig-res1}
}
\end{figure}

\begin{figure}
\bigskip
\caption{ Dependence of Beam Size on Source Flux.
Resolution significantly improves for sources stronger than the
local processing background of 33.7 Jy at 60 {\micron} and
223 Jy at 100 {\micron} in the test field. Results with destriping
(solid lines), and non-destriping (dashed lines) show the IGA destriper
has comparable resolution to standard HIRES, with the most
notable difference along the 100 {\micron} major axis.
\label{fig-res2}
}
\end{figure}

\begin{figure}
\bigskip
\caption{The ratio of (IGA(20) Flux / PSC Flux) vs. PSC Flux shows
         no trend with source strength. Offsets are discussed in text.
         Thirty-five sources are plotted in each wavelength band.
\label{fig-photo}}
\end{figure}


\begin{figure}
\bigskip
\caption{
Illustrations of the Hysteresis Effect. (a). Internal flashes of known
magnitude were used at the starting and ending points of a scanline,
in an effort to determine the responsivity change (dashed line, assumed
responsivity). The true responsivity is shown in the dashed curve, due
to photon induced responsivity enhancement;
(b). Computed fluxes from ascending and descending scans deviate from
the true values; (c). The ratio of fluxes computed from ascending or
descending scans varies with Galactic latitude.
(Adapted with changes from Figure VI.B.2,
{\IRAS} Catalogs and Atlases: Expl. Suppl. 1988.)
}
\label{fig-hyst-exp}
\end{figure}

\begin{figure}
\caption{Residual Hysteresis in the IGA. Left panels show the average intensity versus Galactic latitude at $l$ = 0{\deg}, 60{\deg}, and 120{\deg}. Right panel shows the ratio of ISSA intensities from oppositely directed scans. The hysteresis signature (Fig. \ref{fig-hyst-exp}) is clearly seen near $l$ = 0{\deg}, $b$ = 0{\deg} with an amplitude of about 20\% at 100 {\micron} (upper right panel).
Hysteresis may also be present in the $l$ = 60{\deg} and 300{\deg} graphs but is
below the 5\% level. Other small ($ <$ 5\%) but systematic variations in the
ISSA ratio are likely due to destriping differences. }
\label{fig-hyst-0-120}
\end{figure}

\begin{figure}
\caption{Average Intensity and Ratio of ISSA Intensities from Opposite Scans,
$l$ = 180{\deg}, 240{\deg}, and 300{\deg}.}
\label{fig-hyst-180-300}
\end{figure}

\begin{figure}
\bigskip
\caption{Minimum and Maximum Intensity Ratios vs. Galactic Longitude.
Left: 60 {\micron}; Right: 100 {\micron}. At 100 {\micron} and
within 60{\deg} of the Galactic center, residual hysteresis becomes
larger than systematic differences due to destriping and noise.}
\label{fig-hyst-ratio}
\end{figure}

\begin{figure}
\caption{Demonstration of the Ringing Artifact.
(a). 1st iteration, 60 {\micron};
(b). 20th iteration, 60 {\micron};
(c). 1st iteration, 100 {\micron};
(d). 20th iteration, 100 {\micron}. Ringing is not seen for the 1st iteration
images, but is prominent in the 20th. Field center is at
$l = 75^\circ, b = 1^\circ$; field size is $1.4^\circ$ on each side.
Black is brighter in the images.}
\label{fig-ring}
\end{figure}

\begin{figure}
\bigskip
\caption{Demonstration of the Glitch Artifact. The elongated feature to the
upper-left of the field center is a glitch. (a) and (b) show the glitch
at 1st and 20th iteration respectively. At 20th iteration, the glitch
takes a ``broken-up'' shape. Black is brighter in the images. Field center
is $l = 7^\circ, b = 1^\circ$, field size is $1.4^\circ$ on each side.}
\label{fig-glitch}
\end{figure}

\begin{figure}
\caption{Discontinuity Across One Subfield. 60 {\micron}, 20th iteration,
field center is $l = 48^\circ, b = 1^\circ$, $1.4^\circ$ on each side. The
difference in intensity is approximately 5 MJy/sr. Black is brighter in the images.}
\label{fig-discont}
\end{figure}

\begin{figure}
\caption{ {\IRAS} Galaxy Atlas images of the Galactic plane at 60 {\micron}.
Longitudes 0{\deg} -- 40{\deg} show a variety of
star-forming regions, HII regions, and the diffuse IR emission associated
with the Galactic HI layer.
Each panel covers 11.5{\deg} in longitude and $-1.7${\deg} $< b <$ 1.7{\deg}
in latitude, with logarithmic stretch chosen to emphasize structure.
Black is brighter in the images. }

\label{fig-0-40}
\end{figure}

\begin{figure}
\caption{The Galactic Plane at 60 {\micron}, Longitude 40{\deg} -- 80{\deg}.}
\label{fig-40-80}
\end{figure}

\begin{figure}
\caption{The Galactic Plane at 60 {\micron}, Longitude 280{\deg} -- 320{\deg}.}
\label{fig-280-320}
\end{figure}

\begin{figure}
\caption{The Galactic Plane at 60 {\micron}, Longitude 320{\deg} -- 0{\deg}.}
\label{fig-320-0}
\end{figure}

\clearpage

\begin{figure}
\caption{Example 1.4$^\circ \times 1.4^\circ$ {\IRAS} 
Galaxy Atlas images plus beam sample maps at 60 {\micron} 
near the IC1805 OB cluster. Black is brighter in the images. 
Top left shows the iteration = 1 image at the native {\IRAS}
resolution. There are several sources of diffuse emission and
cirrus filaments in the field. The top right panel containing
the iteration = 20 image illustrates how HIRES 
processing `sharpens' features in the image. Many new discrete 
sources are now visible. Bottom panels: In the beam sample maps
a field of artificial point sources helps to assess the effective 
spatial resolution in the IGA for faint {\IRAS} sources. Bottom left panel 
exhibits the elliptical PSF which is typical in coadded {\IRAS} 
images. Twenty iterations of HIRES processing
(bottom right panel) pulls sources out of the background and 
sharpens the PSF but the point source ringing artifact
appears (see Sec. \ref{ring}). Notice that the PSF
varies over the image. For strong sources, i.e. sources with high source-to-background contrast, the spatial resolution is better 
than indicated by the beam sample maps (see Sec. \ref{res}).}
\label{fig-ica}
\end{figure}

\begin{figure}
\caption{Examples of diagnostic ancillary maps for the field near the
 IC1805 OB cluster. The scanlines are evident in the detector track 
map (upper left) which shows the central positions of the {\IRAS} detector 
samples projected into the image plane. The data coverage map (upper right) is the detector track map (upper left) convolved with the rectangular {\IRAS} detector response profiles. This is the most useful diagnostic map for assessing HIRES image quality. Best HIRES results are obtained for high ($>$25) and uniform coverage. The top border shows several regions of very low coverage (black) which can lead to artifacts (see Sec. \ref{cvg}). The
photometric noise map (bottom left) provides a measure of the internal detector noise. A noisy detector scan will appear as a stripe. The correction factor variance
map (bottom right) measures the fitting error in units of $(S/N)^{-2}$. High CFV values ($>$ 0.1) indicate noisy or unreliable parts of the image along the top and
bottom borders (see Sec. \ref{cvg}).}
\label{fig-icb}
\end{figure}



\clearpage

\begin{table}
\dummytable\label{tmaps}
\end{table}

\begin{table}
\dummytable\label{tphoto}
\end{table}

\begin{table}
\dummytable\label{tphotostat}
\end{table}

\begin{table}
\dummytable\label{tpos}
\end{table}


\begin{table}
\dummytable\label{tmos}
\end{table}


\end{document}